\documentstyle[preprint,aps,epsfig]{revtex}
\begin{document}
\draft
\title{Quasiperiodic time dependent current in
driven superlattices: distorted Poincar\'e
maps and strange attractors}
\author{David S\'anchez$^1$, Gloria Platero$^1$ and Luis
L. Bonilla$^{2,3}$}
\address{$^1$Instituto de Ciencia de Materiales
(CSIC), Cantoblanco, 28049 Madrid, Spain\\
$^2$Escuela Polit{\'e}cnica Superior, Universidad
Carlos III de Madrid, Avenida de la Universidad
20, 28911 Legan{\'e}s, Spain.\\
$^3$Also: Unidad Asociada al Instituto de Ciencia
de Materiales (CSIC)}
\date{\today}
\maketitle
\begin{abstract}
Intriguing routes to chaos have
been experimentally observed in semiconductor
superlattices driven by an ac field. In this
work, a theoretical model of time dependent
transport in ac driven superlattices is
numerically solved. In agreement with experiments,
distorted Poincar\'e maps in the quasiperiodic
regime are found. They indicate the appearance of
very complex attractors and routes to chaos as
the amplitude of the AC signal increases.
Distorted maps are caused by the discrete
well-to-well jump motion of a domain wall during
spiky high-frequency self-sustained oscillations
of the current.
\end{abstract}
\newpage
\narrowtext Nonlinear dynamics of weakly coupled
semiconductor superlattices (SLs) driven by dc
and ac biases has been the research topic of many
experimental and theoretical works
\cite{gra91,pre94,mer95,bon94,agu97,kas97,wac98}.
The nonlinearity manifests itself in many physical
situations as for instance, in their transport
properties under finite DC or AC bias. Under
appropriate DC voltage bias, the current through
the SL displays natural oscillations caused by
creation, motion and annihilation of domain walls
(DWs) in the SL
\cite{bon94,kas97}. The observed oscillation
frequencies range from kHz to GHz. Superimposed
on a smooth current oscillation, there appear
faster current spikes whose frequencies are
typically in the GHz regime \cite{kas97}. Each
spike reflects the well-to-well motion of DWs
causing the self-oscillation, and therefore their
frequency is related to the characteristic
tunneling time. Theory and numerical simulations
show that DW may be charge monopoles or dipoles
\cite{wac98,san99}, although experimental
evidence shows that DW are charge monopoles for
self-oscillations observed in the available
samples
\cite{kas97}. AC driven SLs display a rich
dynamical behavior including quasiperiodic and
chaotic oscillations with nontrivial spatial
structure
\cite{bul95,bul99,zha96,luo98l,luo98b}.

Many studies fix the frequency of the AC drive as
the golden mean number $(1+\sqrt 5)/2\approx
1.618$ times the frequency of the natural
oscillations (i.e., the frequency ratio is an
irrational number hard to approximate by rational
numbers), which is convenient to obtain complex
dynamical behavior \cite{Ott}. In this case, the
system presents a rich power spectrum, a complex
bifurcation diagram and different routes to chaos
including quasiperiodicity, frequency locking,
etc. \cite{bul95,bul99,zha96,luo98l,luo98b}.
First return or Poincar\'e maps are used to
analyze unambiguously the underlying attractors
\cite{luo98b,Ott}. In the quasiperiodic case,
Poincar\'e maps usually consist of smooth loops,
whereas they are a set of discrete points in the
case of frequency locking. More exotic Poincar\'e
maps resembling distorted double loops in the
quasiperiodic case have been experimentally
observed in middle of the second plateau of the
I--V characteristic of a SL
\cite{luo98l,luo98b}. At the onset of this
plateau, Poincar\'e maps are smooth and not
distorted. The origin of distorted maps was not
understood at the time of their observation,
although disorder and sample imperfections were
invoked \cite{luo98l}. Luo et al \cite{luo98b}
showed that a combination of signals with
different frequency was needed in order to
reproduce experimentally observed distorted
double layer Poincar\'e maps. The origin of this
combination was not ascertained in that work.

The aim of the present work is to establish
theoretically that high-frequency current spikes
of the self-oscillations give rise to these
exotic Poincar\'e maps. In turn, current spikes
are due to the well-to-well motion of DW during
each period of the self-oscillations. Thus
distorted Poincar\'e maps reflect DW motion in AC
driven SLs.

Theoretical studies of nonlinear effects in
weakly coupled SL are based in microscopic
modelization of the sequential tunneling current
between adjacent wells \cite{agu97,wac98}. Early
models were discrete drift models with a
phenomenological fitting of tunneling current and
boundary conditions
\cite{bon94}. More sophisticated models have
included the electrostatics at the contacts in a
self-consistent framework\cite{agu97,san99}.
 Depending on SL configuration and contact
doping, we know that undriven self-oscillations
are due to recycling of either monopole DWs or
dipole waves \cite{san99}. In the latter case, a
dipole wave consists of two DWs, one
corresponding to a charge accumulation layer and
another one corresponding to a depletion layer.
Between these DWs there is a high-field region
encompassing several wells, and the whole dipole
wave travels through the SL. Recycling of such
dipole waves gives rise to current
self-oscillations similar to the well-known Gunn
effect in bulk semiconductors \cite{hig92}.

In this paper we have analyzed the tunneling
current through a DC+AC biased SL by means of the
microscopic model described in \cite{agu97,san99}.
We have considered a 50-period SL consisting of
13.3 nm GaAs wells and 2.7 nm AlAs barriers, as
described in \cite{kas97}. Doping in the wells
and in the  contacts are $N_{w} = 2\times
10^{10}$~cm$^{-2}$ and $N_{c}= 2\times
10^{18}$~cm$^{-3}$ respectively. With these
doping values, self-oscillations are due to
recycling of monopole DWs. We choose not to
analyze the original 9/4 sample where the
experiments were performed \cite{luo98l}. In this
sample, the second plateau of the I--V
characteristics ends at a resonant peak
due to $\Gamma - X$ tunneling. In the simpler SL
configuration chosen here, the $X$ point does not
contribute to the peaks at the first and second
plateaus. We analyze the sequential tunneling
current\cite{san99} with an applied voltage,
$V(t) = V_{AC}(t) + V_{DC}$, where $V_{AC}(t)=
V_{AC}\sin (2\pi f_{AC}t)$. The applied AC
frequency $f_{AC}$ has been set to the golden
mean times the natural frequency and is very
small compared with typical energy scales of the
system. Thus the AC potential modifies
adiabatically the potential profile of the SL and
our sequential tunneling model holds
\cite{agu97,san99}.

\begin{figure}[!htp]
\centerline{
\epsfig{file=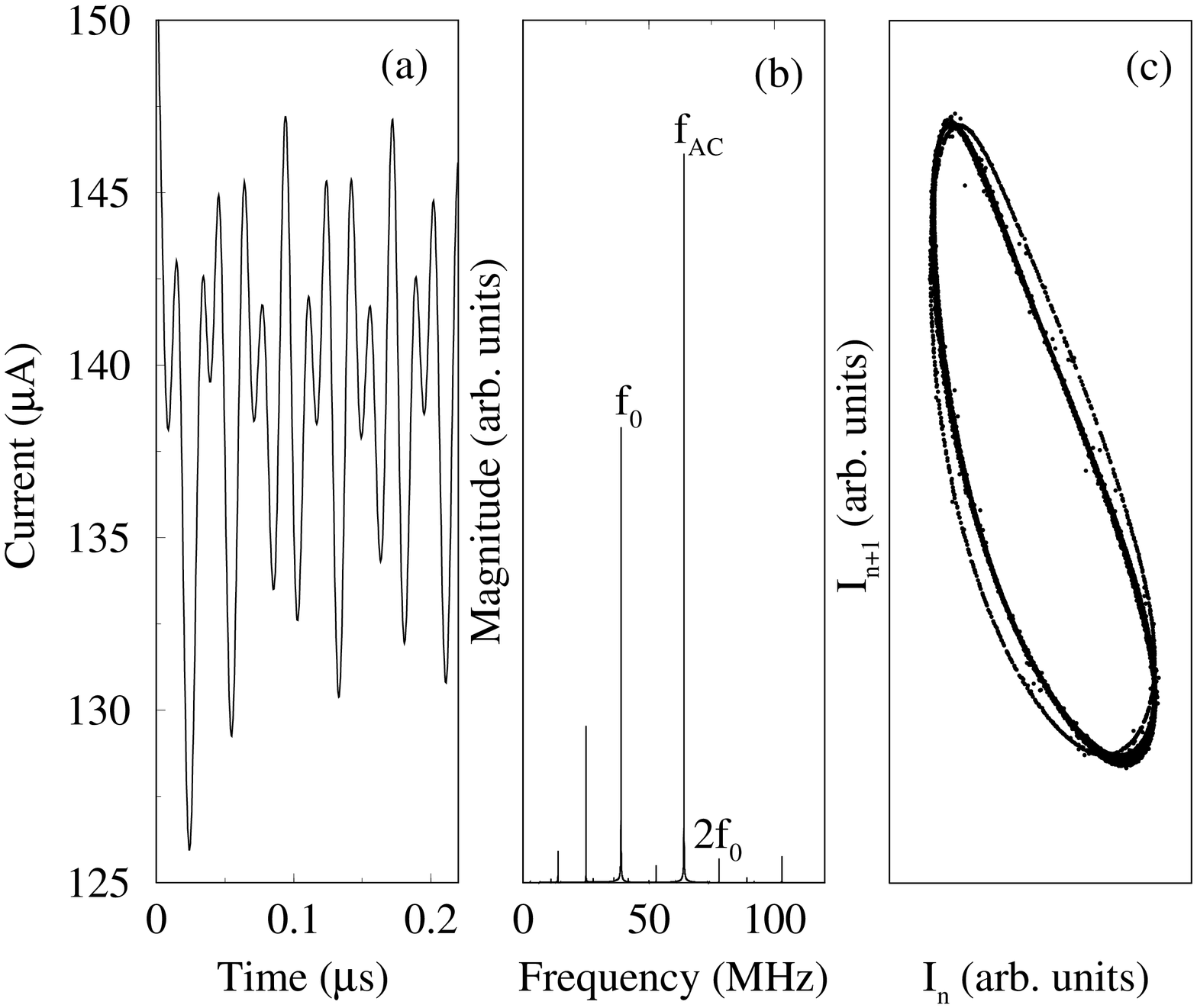,angle=270,width=0.45\textwidth}
}
\caption{(a) I(t) for $V_{DC}=4.2$~V, $f_{0}=39$~MHz, $V_{AC}=19$~mV.
Spikes are not resolved.
(b) Power spectrum.
Notice that higher harmonics of the fundamental frequency are barely formed.
(c) Poincar\'e map, constructed by plotting the current at 
the (n+1)st AC period
versus the current in the preceding period.}
\label{notrans}
\end{figure}

\begin{figure}[!htp]
\centerline{
\epsfig{file=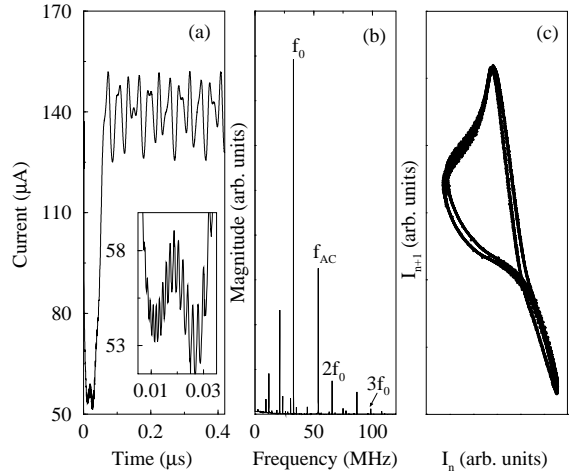,angle=270,width=0.45\textwidth}
}
\caption{(a) I(t) for $V_{DC}=5.1$~V, $f_{0}=33$~MHz.
Spikes are present in the transient regime (see inset).
(b) Power spectrum. Some higher harmonics of $f_{0}$ can be observed.
(c) Poincar\'e map.
Notice that the loop is somewhat distorted due to the presence 
of higher harmonics in (b).}
\label{trans}
\end{figure}

Fig.~\ref{notrans} shows the evolution of the
current through the SL, its Fourier spectrum and
its Poincar\'e map for $V_{DC}=4.2$~V and
$V_{AC}= 19$~mV. These values correspond to the
onset of the second plateau of the I--V
characteristic curve. The current trace of
Fig.~\ref{notrans}(a) is quasiperiodic and does
not present observable superimposed
high-frequency oscillations (spikes).  The
natural oscillation near the onset of the plateau
is caused by monopole recycling very close to the
collector contact. Thus the DW does not move over
many wells and the current trace does not present
an appreciable number of spikes. In the power
spectrum of Fig.~\ref{notrans}(b) there are
contributions coming from the  fundamental
frequency $ f_{0}$=39 MHz, the frequency of the
applied AC field $f_{AC}$, the combination of
both and their higher harmonics. The Poincar\'e
map depicted in Fig.~\ref{notrans}(c) is a smooth
loop with a nontrivial double layer structure
indicating quasiperiodic oscillations. If we
increase the DC bias up to the middle of the
second plateau, $ V_{DC}= 5.1$~V, the undriven
oscillation is due to recycling of monopole DWs
which propagate periodically through part
(approximately $40\%$) of the structure and
disappear at the collector. A similar motion is
observed in the AC driven case. The corresponding
current trace may show high frequency spikes
depending on the chosen initial field profile. A
flat initial field profile gives a spiky current
trace until the latter settles to the stable
oscillation (over which there are no appreciable
spikes); see the earlier part of
Fig.~\ref{trans}(a). The power spectrum and
Poincar\'e maps corresponding to this case and to
the previous one (Figs.~\ref{notrans}(b) and (c))
are markedly different. Fig.~\ref{trans}(b) shows
that the power spectrum of the case with spikes
has a greater harmonic content than that of the
case without spikes (Fig.~\ref{notrans}),
presenting many peaks corresponding to higher
harmonics of the fundamental frequencies (i.e.,
the low fundamental frequency oscillations,
$f_{0}=33$ MHz, and the spikes, $500$~MHz), the
applied frequency and combinations thereof. The
corresponding Poincar\'e map consists of a
distorted loop with a double layer structure
which shows a strong similarity with previous
experiments \cite{luo98l,luo98b}; see
Fig.~\ref{trans}(c). From these numerical
observations, we conclude that distorted
Poincar\'e maps are linked to spiky current
traces, even if such traces change to smooth ones
after a transient.

\begin{figure}[!htp]
\centerline{
\epsfig{file=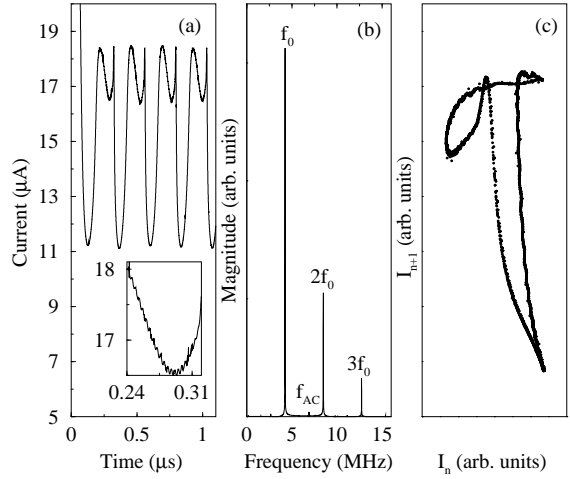,angle=270,width=0.45\textwidth}
}
\caption{(a) I(t) for $V_{DC}=1.5$~V, $f_{0}=4$~MHz, $V_{AC}=2$~mV.
Spikes are superimposed on the current throughout the signal (see inset).
(b) Power spectrum. Higher harmonics of $f_{0}$ contribute with a 
finite amplitude to the power spectrum.
(c) Poincar\'e map. The distortion is greater than in 
Fig.~\ref{trans}(c) (see text).}
\label{dip}
\end{figure}

The previous conclusion may be reinforced if we
change the emitter doping so that undriven
self-oscillations are caused by dipole waves, for
which the current traces are more spiky
\cite{san99}. Thus we carried out numerical
simulations with a smaller contact doping,
$N_{c}= 2\times 10^{16}$~cm$^{-3}$. The frequency
of the natural oscillation is now reduced to 4
MHz and the AC intensity is $2$~mV. The current
at the middle of the first plateau ($V_{DC}$=1.5
V) is shown in Fig.~\ref{dip}(a). At finite time,
$I(t)$ presents dipole-like oscillations and
superimposed finite amplitude spikes. The
Poincar\'e map, Fig.~\ref{dip}(c), is much more
complicated than in the previous case, showing
three well defined distorted loops. Since loops
in the Poincar\'e map are due to combination of
strong enough signals of different frequencies
\cite{luo98b}, the greater strength of the
high-frequency spikes gives rise to the
additional loop structure and higher harmonic
content (Fig.~\ref{dip}(b)).

\begin{figure}[!htp]
\centerline{
\epsfig{file=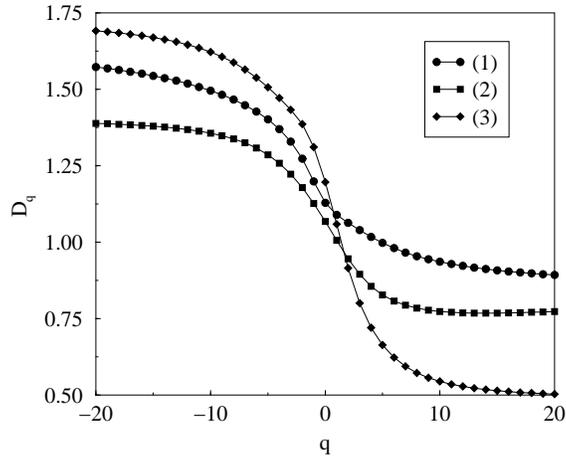,angle=270,width=0.45\textwidth}
}
\caption{Multifractal dimension. The labels (1),
(2) and (3) correspond to the attractors in
Figs.~\ref{notrans}(c),~\ref{trans}(c)~and~\ref{dip}(c),
respectively.}
\label{dq}
\end{figure}

As we mentioned above, the double layer structure
of Poincar\'e maps indicates nontrivial attractors
for the quasiperiodic case. We have  calculated
the multifractal dimensions of the attractors,
$D_q$, for the three cases discussed above; see
Fig.~\ref{dq}. In all cases, they correspond to
strange attractors \cite{bul99,Gre,Dit} whose
$D_q$ presents the knee-like structure typical of
chaotic attractors with multifractal dimensions
\cite{bul99}. A detailed discussion of the
attractor dimensions will be presented elsewhere.

In summary, we have analyzed theoretically the
time dependent current through a AC driven SL. We
have characterized intriguing Poincar\'e maps of
the quasiperiodic oscillations. The strange
attractors which define these Poincar\'e maps
have their physical origin in the complex
dynamics of the domain wall. We have shown that
distorted loops in the Poincar\'e maps are
related to the presence of high frequency spikes
in the current traces. Their frequencies  combine
with the AC frequency and the low natural
frequency to yield a richer power spectrum. Since
spikes are associated to extended motion of the
DW, we can understand why distorted Poincar\'e
maps are observed at the middle of a I--V plateau
and not at its beginning, where the
monopole moves over a too small part of the SL.
The case of natural self-oscillations due to
dipole recycling is different. There the DW of
the dipole are generated at the emitter contact
and move over the whole SL. In such a case,
spikes are more prominent than in the monopole
case, and a higher harmonic content and distorted
Poincar\'e maps appear. Our results may help
explaining recent experimental evidence showing
complex Poincar\'e maps and intriguing routes
to chaos \cite{luo98b}.

{\bf Acknowledgments}.
This work
has been supported by the DGES (Spain) grants
PB98-0142-C04-01 and PB96-0875, by the European
Union TMR contracts ERB FMBX-CT97-0157 and
FMRX-CT98-0180 and by the Community of Madrid,
project 07N/0026/1998.

\end{document}